\documentclass[10pt,twocolumn,superscriptaddress,english,10pt,prl,showpacs,floatfix,aps]{revtex4-1}
\usepackage[latin9]{inputenc}
\usepackage{amsthm}
\usepackage{amsmath}
\usepackage{amssymb}
\usepackage{esint}
\usepackage{graphicx}

\makeatletter

\@ifundefined{textcolor}{}
{%
 \definecolor{BLACK}{gray}{0}
 \definecolor{WHITE}{gray}{1}
 \definecolor{RED}{rgb}{1,0,0}
 \definecolor{GREEN}{rgb}{0,1,0}
 \definecolor{BLUE}{rgb}{0,0,1}
 \definecolor{CYAN}{cmyk}{1,0,0,0}
 \definecolor{MAGENTA}{cmyk}{0,1,0,0}
 \definecolor{YELLOW}{cmyk}{0,0,1,0}
}

\usepackage{braket}
\usepackage[backref=true,
bookmarksnumbered=true,
bookmarks=true,
bookmarksopen=true,
colorlinks=true,
citecolor=blue,
linkcolor=blue,
anchorcolor=green,
urlcolor=blue,unicode=false]{hyperref}

\renewcommand{\v}[1]{\ensuremath{\mathbf{#1}}} 
\newcommand{\gv}[1]{\ensuremath{\mbox{\boldmath$ #1 $}}} 
\newcommand{\abs}[1]{\left| #1 \right|} 
 
\let\baraccent=\= 
\renewcommand{\=}[1]{\stackrel{#1}{=}} 

\DeclareMathOperator{\Tr}{Tr} 
\DeclareMathOperator{\diag}{diag}

\def\lsim{\mathrel{\rlap{\lower4pt\hbox{\hskip1pt$\sim$}}
    \raise1pt\hbox{$<$}}}                
\def\gsim{\mathrel{\rlap{\lower4pt\hbox{\hskip1pt$\sim$}}
    \raise1pt\hbox{$>$}}}                

\makeatother

\usepackage{babel}

\begin{document}

\title{Strong localization of Majorana end states in chains of magnetic adatoms}

\author{Yang Peng}
\affiliation{\mbox{Dahlem Center for Complex Quantum Systems and Fachbereich Physik, Freie Universit{\"a}t Berlin, 14195 Berlin, Germany} }

\author{Falko Pientka}
\affiliation{\mbox{Dahlem Center for Complex Quantum Systems and Fachbereich Physik, Freie Universit{\"a}t Berlin, 14195 Berlin, Germany} }

\author{Leonid I.\ Glazman}
\affiliation{Department of Physics, Yale University, New Haven, CT 06520, USA}

\author{Felix von Oppen}
\affiliation{\mbox{Dahlem Center for Complex Quantum Systems and Fachbereich Physik, Freie Universit{\"a}t Berlin, 14195 Berlin, Germany} }

\begin{abstract}
A recent experiment [Nadj-Perge {\em et al.}, Science {\bf 346}, 602 (2014)] gives possible evidence for Majorana bound states in chains of magnetic adatoms placed on a superconductor. While many features of the observed end states are naturally interpreted in terms of Majoranas, their strong localization remained puzzling. We consider a linear chain of Anderson impurities on a superconductor as a minimal model and treat it largely analytically within mean-field theory. We explore the phase diagram, the subgap excitation spectrum, and the Majorana wavefunctions. Owing to a strong velocity renormalization, the latter are localized on a scale which is parametrically small compared to the coherence length of the host superconductor.
\end{abstract}

\maketitle

{\em Introduction.---}There is currently great interest in Majorana bound states in condensed-matter systems which realize nonabelian quantum statistics \cite{review1,review2} and may have applications in topological quantum information processing \cite{kitaev03}. Several platforms allow one to engineer topological superconducting phases supporting Majorana bound states, based on proximity coupling to $s$-wave superconductors (SC). These include topological insulators \cite{fu08,fu09}, semiconductor quantum wires \cite{lutchyn10,oreg10,alicea11}, and chains of magnetic adatoms \cite{bernevig13,pientka13,loss,braunecker,franz,kim} (see also \cite{beenakker_magnetic,flensberg,morpurgo}). All these proposals are being actively pursued in the laboratory \cite{mourik12,das12,churchill13,rokhinson12,lund,harlingen,yacoby,kouwenhoven,ali}. 

A recent experiment \cite{ali} exhibits signatures of Majorana bound states in chains of Fe atoms placed on a Pb surface. Experiment suggests that the Fe chain orders ferromagnetically. The subgap spectrum is probed by scanning tunneling spectroscopy with both spatial and spectral resolution which shows zero-energy states near the ends of the chains. It is tempting to interpret these as Majorana bound states \cite{ali,li14} as the system combines the three essential ingredients: (i) Proximity-induced superconductivity; (ii) a finite Zeeman splitting due to the exchange field of the ferromagnetic Fe chains; and (iii) Rashba spin-orbit (SO) coupling (presumably from the surface of the Pb substrate). 

However, the observed localization of the end states on the scale of a few adatom sites is puzzling \cite{DasSarma,lee}. The Majorana localization length is typically estimated as $\xi_M=\hbar v_F/\Delta_{\rm top}$, while the coherence length $\xi_0$ of the proximity-providing SC is given by $\xi_0=\hbar v_F/\Delta$. Here, we assume comparable Fermi velocities $v_F$ in the one-dimensional electron system (`wire') and the host SC. At the same time, the induced topological gap $\Delta_{\rm top}$ is smaller than the host gap $\Delta$. Thus, one may expect $\xi_M\gsim \xi_0$. This contrasts with the observation that the localization length of the end states is orders of magnitude smaller than the coherence length of Pb. Here we address this puzzle by modeling the adatoms as a chain of Anderson impurities hybridized with a SC and show that it predicts Majorana localization lengths which are parametrically smaller than $\xi_0$ over wide regions of parameter space.

The physics underlying the topological phase in chains of magnetic adatoms has been discussed in two approaches. One approach  \cite{bernevig13,pientka13,nagaosa,ojanen,kotetes,sau} starts with the subgap Shiba states \cite{yu,shiba,rusinov,rmp} induced by the individual magnetic adatoms. The adatom is described as a classical magnetic moment which is exchange coupled to the electrons in the substrate, but otherwise electronically inert. Such Shiba chains exhibit topological superconducting phases and hence Majorana end states. An alternative approach \cite{ali,li14} starts with exchange-split adatom states. While they are far from the Fermi energy for individual adatoms, hopping between the adatoms of the chain broadens these states into bands. For sufficiently strong hopping, these bands cross the Fermi energy and effectively realize a one-dimensional spin-polarized electron system. In this  band limit, topological superconductivity is induced by proximity, in combination with SO coupling for ferromagnetic chains or helical magnetic order along the chain. As an additional benefit, our model unifies both of these approaches. 

{\em Heuristic considerations.---}We start by discussing conventional proximity coupling of a free-electron `wire' to a bulk $s$-wave SC. The `wire' electrons are described by their Green function $G(k,E) = [E - v_F k \tau_z -\Sigma(k,E)]^{-1}$, where $\tau_i$ denote Pauli matrices in particle-hole space. The self energy $\Sigma$ accounts for the coupling to the SC and takes the familiar form \cite{potter,zyuzin}
\begin{equation}
  \Sigma(k,E) = - \Gamma \frac{E+\Delta\tau_x}{\sqrt{\Delta^2-E^2}}.
  \label{self}
\end{equation}
Here, $\Gamma$ measures the strength of hybridization between wire and SC. Far above the gap, $E\gg\Delta$, the SC behaves as a normal metal and the escape of electrons into the bulk SC is described by $\Sigma \simeq i\Gamma$. For subgap energies, electrons enter the SC only virtually and $\Sigma$ becomes real. 

For definiteness, consider energies far below the bulk gap, $E\ll\Delta$. Then, we can approximate $\Sigma\simeq -\frac{\Gamma}{\Delta}E -\Gamma \tau_x)$, and $G(k,E)\simeq Z[E - Z v_F k\tau_z -Z \Gamma\tau_x]^{-1}$ with a renormalized quasiparticle weight $Z = [1+\Gamma/\Delta]^{-1}$, which describes the shift of the electrons' spectral weight from the wire into the SC. The quasiparticle weight ensures \cite{review1} that the induced $s$-wave gap (described by the pairing term $\propto\tau_x$) interpolates between the hybridization strength $\Gamma$ at weak hybridization, $\Gamma\ll\Delta$, and the host gap $\Delta$ at strong hybridization, $\Gamma\gg\Delta$. It also renormalizes the Fermi velocity $v_F\to \tilde{v}_F = Z v_F$ which controls the coherence length of the induced superconductivity in the wire. Physically, the fraction of time an excitation spends in the wire is suppressed by $Z$, which reduces the effective velocity to $Z v_F$.

In adatom chains, the SO coupling in the SC allows for induced $p$-wave pairing while the strong on-site repulsion and resulting spin polarization suppress $s$-wave correlations. Thus, the induced gap $\Delta_{\rm top}=\alpha\Delta$ is now $p$-wave and controlled by the (dimensionless) SO strength $\alpha$. At the same time, it is natural to assume that the hybridization $\Gamma$ modifies {\it single-particle} properties as before and the renormalization of $v_F$ remains operative. This predicts a Majorana localization length
\begin{equation}
     \xi_M={\hbar \tilde{v}_F}/{\Delta_{\rm top}}=Z {\hbar v_F}/{\Delta_{\rm top}}.
    \label{Majorana}
\end{equation}
For Fe adatoms in Pb, the hybridization is controlled by atomic scales so that $\Gamma \sim 1$eV \cite{ali}. When compared to the host gap $\Delta \sim 10$K, we find $Z\sim 10^{-3}$! This can dramatically suppress $\xi_M$ relative to the host coherence length $\xi_0 \sim \hbar v_F/\Delta$ ($\simeq 100$nm for Pb). In fact, $\xi_M\sim \xi_0 (\Delta/\Gamma)(\Delta/\Delta_{\rm top})$ so that for $\alpha=\Delta_{\rm top}/\Delta\sim0.1$, the Majorana localization length $\xi_M$ becomes of the order of the spacing between adatoms, as observed in Ref.\ \cite{ali}.

{\em Model.---}We now show that these heuristic arguments are borne out in a microsopic model. We model the system as a linear chain of Anderson impurities placed in an $s$-wave SC. Each adatom hosts a spin-degenerate level of energy $\epsilon_d$ with on-site Hubbard repulsion $U$, representing the Fe $d$-levels. We include nearest-neighbor hopping of strength $w$ between these $d$-levels as well as hybridization of strength $t$ between the $d$-levels and the SC. The model Hamiltonian 
\begin{equation}
  {\cal H} = {\cal H}_d + {\cal H}_s + {\cal H}_T
\label{model}
\end{equation}
contains the BCS Hamiltonian ${\cal H}_s$ of the SC \cite{Delta}, the chain of $d$-levels
\begin{eqnarray}
   {\cal H}_d &=& \sum_{j,\sigma} (\epsilon_d -\mu) d_{j,\sigma}^\dagger d_{j,\sigma} + U \sum_j n_{j\uparrow}^\dagger n_{j\downarrow} \nonumber\\
    && -w \sum_{j,\sigma} [d_{j+1,\sigma}^\dagger d_{j,\sigma} + d_{j,\sigma}^\dagger d_{j+1,\sigma}], 
\end{eqnarray}
and their hybridization with the SC,
\begin{equation}
  {\cal H}_T = -t \sum_{j,\sigma} [\psi_{\sigma}^\dagger({\bf R}_j) d_{j,\sigma} + d_{j,\sigma}^\dagger \psi_{\sigma}({\bf R}_j)].
\end{equation}
Here, $d_{j,\sigma}$ annihilates a spin-$\sigma$ electron in the $d$-level at site ${\bf R}_j=ja {\bf \hat x}$ of the chain, $n_{j,\sigma}= d_{j,\sigma}^\dagger d_{j,\sigma}$, and $\psi_{\sigma}({\bf r})$ annihilates electrons at position ${\bf r}$ (taken as continuous) in the SC. 

The model in Eq.\ (\ref{model}) generalizes the Shiba chain model considered in \cite{bernevig13} and \cite{pientka13}. It reduces to the Shiba chain in the limit of negligible spin fluctuations and  weak intersite hopping $w$. Here, we include the hopping and the ensuing electronic dynamics of the magnetic adatoms within a mean-field treatment of the Hubbard term \cite{anderson,shiba1973},
\begin{eqnarray}
  U n_{j\uparrow}^\dagger n_{j\downarrow} \to \frac{U}{2}\sum_\sigma [\langle n_j \rangle n_{j,\sigma} + \langle m_j\rangle \sigma n_{j,\sigma}],
\end{eqnarray} 
where we defined the occupation $n_j = \sum_\sigma n_{j,\sigma}$ and the site polarization $m_j=n_{j,\uparrow}-n_{j,\downarrow}$. The first term merely renormalizes $\epsilon_d$ and will be absorbed in the following. The second term introduces a local exchange coupling in the adatom orbitals. 

As we are predominantly interested in the localization of the Majorana modes, we do not aim at a self-consistent solution of the mean-field theory. Instead, we accept the formation of a spontaneous moment as experimental fact and explore its consequences. In experiment, the moments order ferromagnetically along the chain. In this case, topological superconductivity requires Rashba SO coupling in the substrate SC \cite{ali,li14,duckheim,chung}. For analytical tractability, we assume instead that the moments develop helical order ${\bf S}_j=(\sin\theta\cos\phi_j, \sin\theta \sin\phi_j, \cos\theta)$ with $\phi_j=2k_hja$ and $\theta=\pi/2$. We emphasize that the model with helical order can be mapped to a ferromagnetic model with SO coupling in both, the adatom $d$-band and the substrate SC. Strictly speaking, the substrate SO coupling generated by the mapping differs from conventional Rashba coupling, but it does include the specific term that allows for proximity-induced $p$-wave pairing. The mapping is effected by the unitary transformation $d_j \to e^{-ik_hja\sigma_z}d_j$ and $\psi({\bf r})\to e^{-ik_hx\sigma_z}\psi({\bf r})$ which rotates the spin basis along the direction of the local impurity moments \cite{loss0,supp}.  

{\em Excitation spectrum and phase diagram.---}In mean-field theory, we can describe the system equivalently by the corresponding Bogoliubov-de Gennes Hamiltonian $H=H_d + H_s + H_T$ (after the above-mentioned unitary transformation) and consider the Green function $G=(E-H)^{-1}$. In view of the local nature of the hybridization $H_T$, we can write a closed set of equations for the restricted Green function $g_{ij} = G({\bf R}_i,{\bf R}_j)$ defined at the sites of the adatoms, 
\begin{align}
\left(\begin{array}{cc}
(g_0^{\rm ss})^{-1} & t\tau_{z}\\
t\tau_{z} & E-H_d
\end{array}\right)g=1.
\label{eq:GF_eq}
\end{align}
We use the Pauli matrices $\tau_i$ ($\sigma_i$) in particle-hole (spin) space. The bare Green function of the SC restricted to the adatom sites and subgap energies is readily obtained within BCS theory
(see \cite{supp} for more details),
\begin{eqnarray}
g_{0,ij}^{\rm ss}(E)&=&-\pi\nu_0\exp({-ik_hx_{ij}\sigma_z})
\label{eq:gss}\\
&\times&\left\{\frac{E+\Delta\tau_x}{\sqrt{\Delta^2-E^2}}{\rm Im}f(r_{ij})+\tau_z{\rm Re} f(r_{ij})\right\},
\nonumber
\end{eqnarray}
where $\nu_0$ is the normal density of state at the Fermi level, $f(r) = e^{ik_F r-r/\xi_E}/k_F r$ and $\xi_E = \hbar v_F/\sqrt{\Delta^2 - E^2}$. Eq.\ (\ref{eq:gss}) is valid for $i\neq j$, but also applies to $i=j$ when dropping the ${\rm Re}f$ term. Here, the factor $\exp({-ik_hx_{ij}\sigma_z})$ is induced by the unitary transformation. 

\begin{figure}
\includegraphics[width=0.43\textwidth]{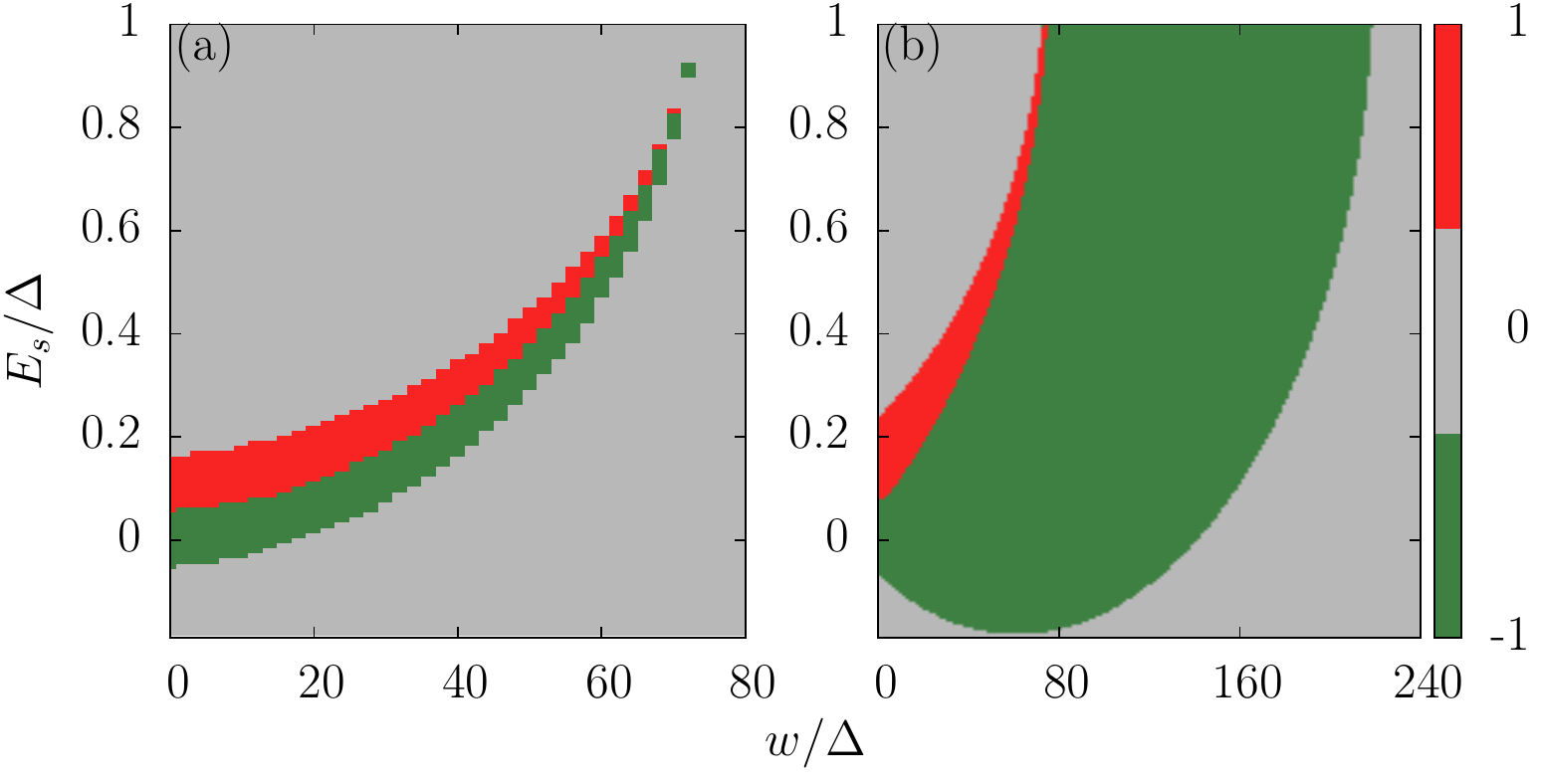}
\caption{\label{PhaseDiagram} Representative phase diagrams for the adatom chain as a function of the Shiba state energy $E_s$ of an individual impurity and the hopping amplitude $w$ between $d$-levels. The colors indicate the topological index (grey: topologically trivial; red/green: topological phase with index $\pm1$). We chose $E_{d,\downarrow}=100\Delta$, $k_F a=4.3\pi$, $k_h a=0.26\pi$, and $\xi_0/a=\infty$. The panels correspond to (a) symmetric adatom $d$-bands ($E_{d,\uparrow}=-100\Delta$) and (b) asymmetric adatom $d$-bands ($E_{d,\uparrow}=-300\Delta$). Here, $E_{d,\sigma} = \epsilon_d - \sigma U\langle m\rangle /2$.}
\end{figure}

The subgap excitation spectrum may then be obtained from the poles of $g^{\rm ss} = g_0^{\rm ss} [1-\Sigma g_0^{\rm ss}]^{-1}$ where we defined the self energy $\Sigma = tg_0^{\rm dd}t=t(E-H_d)^{-1}t$. As $g_0^{\rm ss}$ has no poles at subgap energies, this yields the condition ${\rm det}[1-\Sigma g_0^{\rm ss}] =0$. In (lattice) momentum representation, the determinant involves a $4\times4$ matrix with \cite{supp}
\begin{equation}
  g^{\rm ss}_0(k,E) =\pi\nu_0 \left\{ \frac{E+\Delta \tau_x}{\sqrt{\Delta^2-E^2}} L_i^{\sigma_z}(k,E) + \tau_z  L_r^{\sigma_z}(k,E) \right\}. 
\end{equation}
Here, $L_r^{\sigma_z}$ and $L_i^{\sigma_z}$ are real and imaginary parts of the function $L^{\sigma_z} = F(k+k_h\sigma_z)-i$, respectively, with $F(k)=\frac{1}{k_Fa}\ln\{1-e^{i(k_F+k)a -a/\xi_E}\} + (k\leftrightarrow -k)$ \cite{L}. Computing the dispersions and identifying phase boundaries by the closing of the gap, we first obtain representative phase diagrams of the adatom chain as shown in Fig.\ \ref{PhaseDiagram}. 

These phase diagrams plot the topological (BDI \cite{Zindex}) index and make the interpolation between the band and Shiba limits explicit. The Shiba limit corresponds to weak hopping $w$ between $d$-levels. Here, topological superconductivity requires deep Shiba states so that the Shiba bands cross the chemical potential at the center of the host gap \cite{pientka13}. The band limit corresponds to weak hybridization $\Gamma = \pi\nu_0 t^2$ and thus Shiba states with energies $E_s$ near $\Delta$ \cite{shiba1973,supp}. Then, topological superconductivity requires that one spin-polarized $d$-band crosses the Fermi energy. The range over which this happens depends on the asymmetry of the bare exchange-split adatom states $E_{d,\sigma} = \epsilon_d - \sigma U\langle m\rangle /2$ about the chemical potential (set to $\mu=0$). Fig.\ \ref{PhaseDiagram}(a) shows the symmetric case $E_{d,\uparrow}=-E_{d,\downarrow}$. There is only a narrow topological interval in $w$ for small $\Gamma$ ($E_s\simeq \Delta$) because despite the large exchange splitting of the $d$-levels, the spin-split $d$-bands cross $\mu$ at the same hopping strength $w$. As the asymmetry between $E_{d,\uparrow}$ and $E_{d,\downarrow}$ about $\mu$ increases, the $d$-bands cross $\mu$ at different values of $w$, and the adatom states are perfectly spin polarized at the chemical potential over a substantial region in $w$, cf.\ Fig.\ \ref{PhaseDiagram}(b). 

\begin{figure}
    \includegraphics[width=0.49\textwidth]{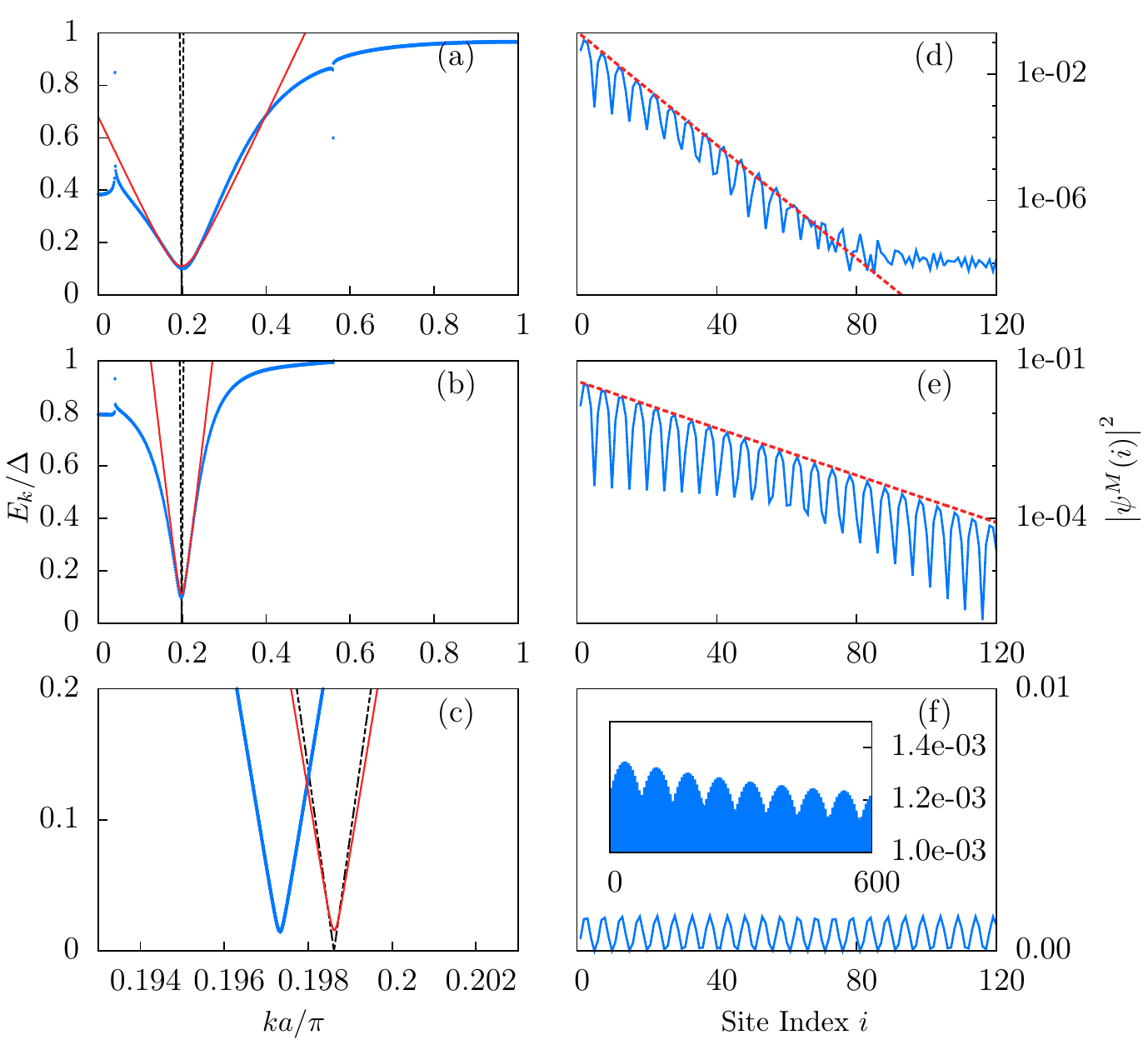}
    \caption{\label{fig:bandmwf} Excitation spectra $E_k$ for $ka/\pi \in [0,1]$ and (a) $\Gamma=64\Delta$, (b) $\Gamma = 16 \Delta$, and (c) $\Gamma = 0.16\Delta $. We chose $k_Fa=4.3 \pi$, $k_h a=0.26\pi$, $E_{d,\downarrow}=100 \Delta$, $E_{d,\uparrow}=-19900\Delta$, $w=90\Delta$, and $\xi_0/a=\infty$. The dashed lines are subgap dispersions of the impurity chain without coupling to the SC. The blue curves are exact dispersions. The red curves are calculated using Eq.\ (\ref{gg}) for $\Gamma\gg \Delta$ and (\ref{ll}) for $\Gamma\ll\Delta$. Notice that the horizontal axis in (c) is restricted to a very narrow range and that the deviation between the red and blue curves is indeed small. Panels (d), (e), and (f) show Majorana wavefunctions $\abs{\psi^M(i)}^2$ (blue lines) obtained for a finite chain of length $L=1500a$. Only the first $120$ sites $i$ are shown. (d) and (e) are plotted on a logarithmic scale and the red dashed lines are fits using Eq.\ (\ref{Majorana}) for the Majorana localization length. (f) is plotted on a linear scale. Inset: Decay over the first 600 sites.}
\end{figure}

For fully analytical results, we consider the limit of strong asymmetry with $E_{d,\uparrow}\to -\infty$ at fixed $E_{d,\downarrow}$. In this limit, only the spin-down band $E_{d} = E_{d,\downarrow} - w \sum_\pm\cos (k\pm k_h)a$ of the $d$-levels is relevant.
A detailed but straightforward calculation \cite{supp} now shows that the condition ${\rm det}(1-\Sigma g^{\rm ss}_0)=0$ can be reduced to the determinant of a $2\times2$-matrix and written in the form
\begin{eqnarray}
  &&(\Delta^2-E^2)[E_d+\Gamma L_r]^2  \nonumber\\
    &&\,\,\,\, - E^2 [\sqrt{\Delta^2-E^2}-\Gamma L_i]^2 + \Gamma^2 \Delta^2 (\delta L_i)^2 = 0 .
\label{central}
\end{eqnarray}
Here, we introduced the shorthand notations $L_{r/i} = (L_{r/i}^+ + L_{r,i}^-)/2$ and $\delta L_{i} = (L_{i}^+ - L_{i}^-)/2$. Eq.\ (\ref{central}) is an implicit equation for the subgap excitation spectrum $E_k$ of the adatom chain in the strongly asymmetric limit. [Note that we have suppressed all $k$ labels in Eq.\ (\ref{central}).]  

In the limits $\Gamma\ll\Delta$ and $\Gamma\gg\Delta$, Eq.\ (\ref{central}) gives explicit analytical expressions for the excitation spectrum throughout the entire Brillouin zone. These are obtained by keeping only the respective dominant term in the second square brackets on the left hand side, in excellent agreement with the full Green-function solution in Fig.\ \ref{fig:bandmwf}(a)-(c). We note that there is a single subgap state for every lattice momentum $k$, i.e., there is one subgap state per adatom, as in the Shiba limit (small $w$). 

{\em Majorana wavefunction.---}Eq.\ (\ref{central}) also encapsulates the localization of the Majorana wavefunctions. In the Shiba limit of small $w$, the Majorana localization was addressed previously \cite{pientka14}. Here, we focus on the band limit of large $w$ where the spin-down $d$-band $E_d$ crosses the Fermi energy of the SC, as is presumably the case in the experiment \cite{ali,li14}. $E_d$ crosses $\mu=0$ at momenta $k_0$, so that $E_d \simeq v_F (k-k_0)$, where $v_F$ is the Fermi velocity of the $d$-band at the chemical potential of the SC. Similarly, $E_d+\Gamma L_r \simeq v_F(k-k_0)$ where we simply absorb the parametrically small shifts in $v_F$ and $k_0$ due to $\Gamma L_r$ into their definitions.

The decay of the Majorana wavefunction is controlled by the behavior of the dispersion near the minimal gap at $k_0$. Assuming that the pitch of the spin helix (or, equivalently, the strength of SO coupling) is not too large, this topological gap will be small compared to the gap $\Delta$ of the superconducting host. Then, $E$ is small compared to $\Delta$ in the relevant region and Eq.\ (\ref{central}) simplifies significantly. Consider first the limit of weak hybridization $\Gamma\ll\Delta$. In this limit, Eq.\ (\ref{central}) reduces to 
\begin{equation}
  E_k = \pm\sqrt{[v_F(k-k_0)]^2 + \Gamma^2(\delta L_i)^2},
  \label{ll}
\end{equation}
where $\delta L_i$ should be evaluated at $k_0$. We identify the topological gap $\Delta_{\rm top} = \Gamma(\delta L_i)_{k=k_0}$ which is small compared to $\Delta$. The Majorana wavefunction is expected to decay on the characteristic length scale of this dispersion, i.e., we find the Majorana localization length $\xi_M = \hbar v_F/\Delta_{\rm top}$, consistent with the heuristic argument above for $\Gamma\ll\Delta$. For the numerical parameters of Fig.\ \ref{fig:bandmwf}(c), $\xi_M$ is larger than the length of the chain, making a direct comparison impossible. 

The experiment is in the limit of large hybridization $\Gamma\gg\Delta$, where Eq.\ (\ref{central}) predicts a low-energy dispersion 
\begin{equation}
  E_k = \pm\sqrt{[(\Delta/\Gamma L_i) v_F(k-k_0)]^2 + [\Delta (\delta L_i/L_i)]^2}.
\label{gg}
\end{equation}
In this limit, the induced gap $\Delta_{\rm top} = \Delta (\delta L_i/L_i)_{k=k_0}$ is independent of $\Gamma$ and saturates to a value which is smaller than $\Delta$ by a factor measuring the effective strength of the SO coupling. The strong hybridization with the SC also induces a dramatic downward renormalization of the Fermi velocity of the excitations, $v_F\to\tilde v_F = (\Delta/\Gamma L_i)v_F$. These features are in excellent agreement with the numerical subgap spectra as shown in Figs.\ \ref{fig:bandmwf}(a) and (b) and vindicate our introductory heuristic arguments. Indeed, Eq.\ (\ref{gg}) predicts a Majorana localization length $\xi_M= \hbar v_F /(\Gamma \delta L_i)$,  which coincides with Eq.\ (\ref{Majorana}) from heuristic consideration. We see that $\xi_M$ is independent of the host gap $\Delta$ and controlled instead by the hybridization $\Gamma$. This result is in excellent agreement with numerical Majorana wavefunctions for $\Gamma\gg\Delta$, see Figs.\ \ref{fig:bandmwf}(d) and (e). 

The topological gaps in Eqs.\ (\ref{ll}) and (\ref{gg}) are both enabled explicitly by the SO coupling in the substrate which enters via the $L$-factors in $g_0^{ss}$. In contrast, the SO coupling in the $d$-band is fully ineffective due to the strong spin polarization. Parametrically, one finds $\delta L_i\simeq \delta L_i/L_i\simeq k_h/k_F$ in the limit $k_Fa\gg 1$. 

Notice that Eqs.~(\ref{gg}) and (\ref{Majorana}) require the condition $\Gamma\ll v_F/a$. This condition ensures that the in-band propagation between adjacent sites, taking time $\tau \sim a/\tilde v_F$, is faster than hopping via the host SC, taking time $(\Delta L_r)^{-1}$  \cite{pientka13}. Then, the $k=k_0$ minimum described by Eq.\ (\ref{gg}) dominates over the additional features of the quasiparticle spectrum associated with logarithmic divergencies in $L_r$. They induce power-law tails in the Majorana wavefunctions [cf.\ Fig.\ \ref{fig:bandmwf}(d)] which become correspondingly more pronounced as $\Gamma$ increases. 

\begin{figure}
    \includegraphics[width=0.49\textwidth]{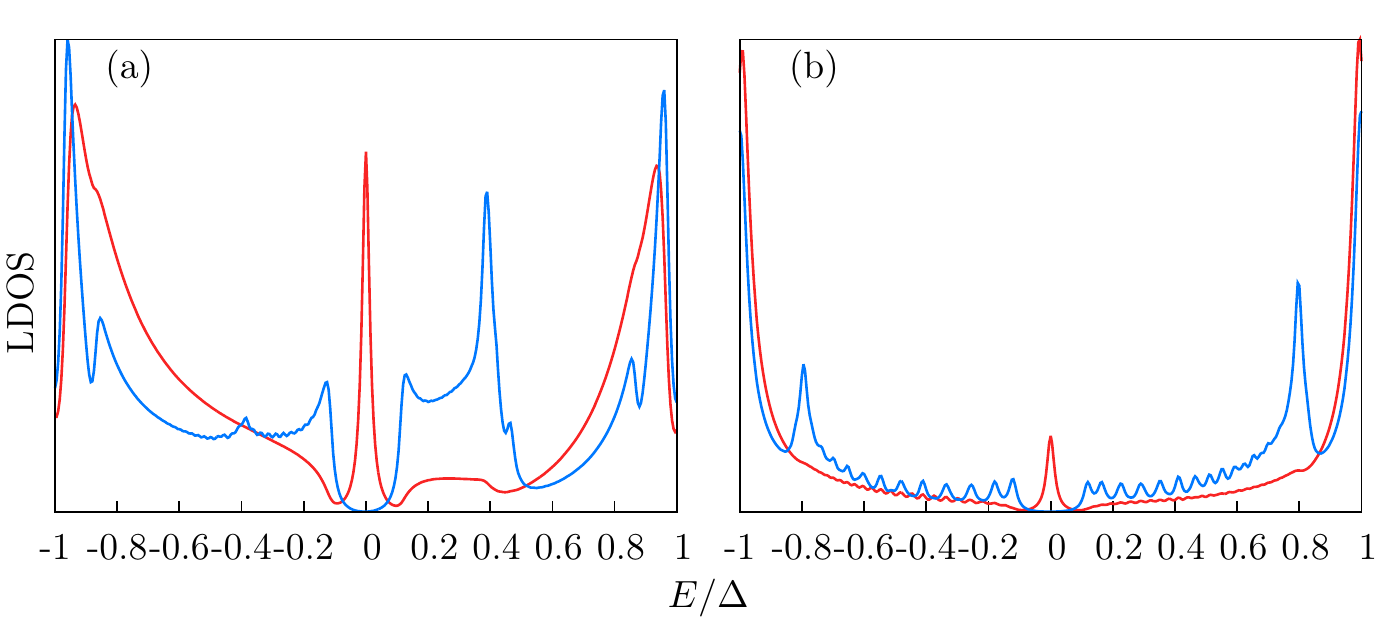}
    \caption{\label{fig:ldos} Local density of states of particle excitations, computed in the center (blue) and at the end (red), for a chain of length $L=300a$ and hybridizations (a) $\Gamma=64\Delta$ and (b) $\Gamma=16\Delta$. Other parameters are as in Fig.\ \ref{fig:bandmwf}. }
\end{figure}

{\em Local density of states.---}We have also numerically computed \cite{supp} the local density of states of the adatom chain, see Fig.\ \ref{fig:ldos}. The zero-bias peak grows more pronounced with increasing $\Gamma$, reflecting the stronger localization of the Majorana wavefunction. In addition to the zero-energy Majorana peak, one discerns additional peaks at finite energies which arise from van Hove singularities in the subgap Shiba band and which approach the center of the gap as the hybridization $\Gamma$ increases. The zero-energy features and their strong localization as well as the van Hove peaks are consistent with the experimental observations \cite{ali}.  

{\em Acknowledgments.---} We thank Ali Yazdani, Andrei Bernevig, Titus Neupert, Piet Brouwer, and Allan MacDonald for stimulating discussions. We acknowledge financial support by the Helmholtz Virtual Institute ``New states of matter and their excitations," SPP1285 and SPP1666 of the Deutsche Forschungsgemeinschaft, the Humboldt Foundation, NSF DMR Grant 1206612, and ONR Grant Q00704. We are grateful to the Aspen Center for Physics, supported by NSF Grant No.\ PHYS-106629, for hospitality while this line of work was initiated.

\newpage
\setcounter{equation}{0}
\begin{widetext}
\section*{Supplementary material}
\section{Unitary transformation of Hamiltonian with helical order}

The mean-field Hamiltonian for a chain of Anderson impurities coupled to an $s$-wave superconductor can be written as
\begin{equation}
    \mathcal{H}=\mathcal{H}_{s}+\mathcal{H}_{d}+\mathcal{H}_{T},
\end{equation}
with
\begin{equation}
    \mathcal{H}_{d}=\sum_{j,\sigma}(\epsilon_d-\mu)\tilde{d}_{j,\sigma}^{\dagger}\tilde{d}_{j,\sigma}
    -w\sum_{j,\sigma}\left(\tilde{d}_{j,\sigma}^{\dagger}\tilde{d}_{j+1,\sigma}
    +\tilde{d}_{j+1,\sigma}^{\dagger}\tilde{d}_{j,\sigma}\right)
    -K\sum_{j,\sigma,\sigma'} \v{S}_{j} \tilde{d}_{j,\sigma}^\dagger \gv{\sigma}_{\sigma,\sigma'}\tilde{d}_{j,\sigma'},
\end{equation}
where the exchange term $K\abs{\v{S}_{j}}=U/2$ arises from a mean-field treatment of the local Hubbard interaction as described in the main text. 
Using the Nambu spinor notation $\tilde{d}_{j}=(\tilde{d}_{j\uparrow},\tilde{d}_{j\downarrow},\tilde{d}_{j\downarrow}^{\dagger},-\tilde{d}_{j\uparrow}^{\dagger})^T$, 
we can write down its Bogoliubov-de Gennes Hamiltonian 
\begin{gather}
    \mathcal{H}_{d}=\frac{1}{2}\sum_{ij}\tilde{d}_{i}^{\dagger}\tilde{H}_{d}^{ij}\tilde{d}_{j} \\
    \tilde{H}_d^{ij}= \left[(\epsilon_{d}-\mu)\tau_{z}-K\mathbf{S}_{j}\cdot\gv{\sigma}\right]\delta_{ij}
    -w\tau_z(\delta_{i,j-1}+\delta_{i,j+1}),
\end{gather}
where the $\tau_{i}$ with $i=x,y,z$ are Pauli matrices in particle-hole space. The s-wave superconductor is modeled by the BCS Hamiltonian
\begin{gather}
    \mathcal{H}_{s}=\frac{1}{2}\int d^{3}r\,\tilde{\psi}^{\dagger}(\mathbf{r})\tilde{H}_{s}\tilde{\psi}(\mathbf{r})  \\
    \tilde{H}_{s}=\xi_{{\mathbf{p}}}\tau_{z}+\Delta\tau_{x} \\
    \xi_{{\mathbf{p}}}=\frac{\mathbf{{p}}^{2}}{2m}-\mu,
\end{gather}
where $\tilde{\psi}(\mathbf{r})=(\tilde{\psi}_{\uparrow}(\mathbf{r}),\tilde{\psi}_{\downarrow}(\mathbf{r}), \tilde{\psi}_{\downarrow}^{\dagger} (\mathbf{r}),-\tilde{\psi}_{\uparrow}^{\dagger}(\mathbf{r}))^T$, 
and $\Delta$ is the superconducting order parameter. The hybridization between the magnetic adatoms and the superconductor in particle-hole space is given by
\begin{equation}
    \mathcal{H}_{T}=-\frac{t}{2}\sum_{j}\left(\tilde{\psi}^{\dagger}(\mathbf{R}_{j})\tau_{z}\tilde{d}_{j}+h.c.\right),
\end{equation}
where $\v{R}_j=ja{\bf \hat{x}}$ denotes the position of the $j$th magnetic adatom, $t$ the hybridization strength, and $a$ the lattice spacing along the chain (i.e., the $x$) direction. 

We assume  a spin helix configuration 
\begin{equation}
    \v{S}_j = (\sin\theta_j \cos\phi_j,\sin\theta_j\sin\phi_j,\cos\theta_j)
\end{equation}
with $\theta_j=\theta$ and $\phi_j=2 k_h ja$, i.e., the spin rotates about the $z$-axis with wavevector $2k_h$ and opening angle $\theta$. This is equivalent to a ferromagnetic configuration with a particular type of spin-orbit coupling via a unitary transformation. Explicitly, we transform  $\psi(\mathbf{r})=e^{ik_{h}x\sigma_{z}}\tilde{\psi}(\mathbf{r})$ and $d_{j}=e^{ik_{h}ja\sigma_{z}}\tilde{d}_{j}$, which rotates the local spin quantization axis along the local direction of the magnetic moment, or equivalently, maps the system on a ferromagnetic configuration.
For the superconductor, this yields 
\begin{gather}
    \mathcal{H}_{s}=\frac{1}{2}\int d^{3}r\psi^{\dagger}(\mathbf{r})H_{s}\psi(\mathbf{r})\\
    H_{s}=\left[\frac{(\v{p}+k_h\sigma_z\hat{x})^2}{2m}\right]\tau_{z}+\Delta\tau_{x}\label{Hs} .
\end{gather}
The Hamiltonian for the chain of magnetic adatoms transforms into
\begin{gather}
    \mathcal{H}_{d}  =\frac{1}{2}\sum_{ij}d_{i}^{\dagger}H_d^{ij}d_{j} \\
    H_{d}^{ij}= \left[(\epsilon_{d}-\mu)\delta_{ij}-W_{ij}\right]\tau_{z}-K\v{S}\cdot\gv{\sigma}\delta_{ij} \label{Hd}
\end{gather}
with $W_{ij}=-we^{-ik_ha\sigma_z}$ if $i=j+1$, $W_{ij}=-we^{ik_ha\sigma_z}$ if $i=j-1$, and zero otherwise. Finally we have the hybridization term
 \begin{equation}
     \mathcal{H}_{t}=-\frac{t}{2}\sum_{j}\left(\psi^{\dagger}(\mathbf{R}_{j})\tau_{z}d_{j}+h.c.\right),
 \end{equation}
which is invariant under this transformation.

We observe that in the transformed Hamiltonian, the helix wavevector plays the role of the strength of (a particular type of) spin-orbit coupling. There is spin-orbit coupling in both the wire and the superconductor. It can be seen from the results of the main text that it is predominantly the spin-orbit coupling in the superconductor that is operative in inducing a $p$-wave gap in the excitation spectrum. This is true as long as the exchange splitting of the $d$-bands is large compared to the effective spin-orbit strength. 

The angle between the exchange splitting and the spin-orbit field depends on the opening angle $\theta$ of the spin helix. The optimal situation for topological superconductivity is when exchange splitting and spin-orbit field are orthogonal to one another. This happens for a planar spin helix $\theta=\pi/2$ which is what we consider in the main text and in the following. Explicitly, for this choice the exchange field is along the $x$-direction, $K\v{S}\cdot\gv{\sigma}=KS\sigma_x$, with $S=\abs{\v{S}}$, while the spin-orbit field is along the $z$-direction. 

Note that the spin-orbit coupling contains only the momentum along the chain. This is different from a conventional Rashba coupling where momenta along both directions of the surface would appear. Presumably, this has mostly quantitative consequences as it is the momentum along the chain which is essential for allowing induced $p$-wave pairing in the adatom band. 

We finally note that one may also want to include pairing correlations $\langle d_{i\uparrow}d_{j\downarrow}\rangle$ in the mean field approximation for the Hubbard interaction on the adatom sites. We have neglected them as we assume a large exchange splitting of the $d$-bands which should strongly suppress any influence of these additional pairing correlations. 

\section{Shiba state energy $E_s$ for an individual impurity}

A single Anderson impurity hybridized with a BCS superconductor induces a subgap state called Shiba state. At mean-field level (as treated in this paper) and large spin splitting, the Shiba state energy is given by \cite{shiba1973}
\begin{equation}
    E_s =\frac{E_{d,\uparrow}E_{d,\downarrow}-\Gamma^2}{\sqrt{(E_{d,\uparrow}E_{d,\downarrow}-\Gamma^2)^2-\Gamma^2(E_{d,\uparrow}-E_{d,\downarrow})^2}}.
\end{equation}

\section{Derivation of the lattice Green function}
 
We write the Hamiltonian of the total system as
 \begin{equation}
     H=\left(\begin{array}{cc}
         H_{s} & H_{t}\\
         H_{t}^{\dagger} & H_{d}
     \end{array}\right)=H_{0}+H_{T}, \label{eq.H}
 \end{equation}
 where $H_s$ and $H_d$ were defined in Eqs.\ (\ref{Hs}) and (\ref{Hd}), $H_0=\diag(H_s,H_d)$ contains the diagonal terms, and
 $H_T$ describes the hybridization on the off-diagonal 
     \begin{gather}
         H_t^{\v{r},j} = -t\delta(\v{r}-\v{R}_j) \tau_z. \label{Ht}
     \end{gather}
     Define the wavefunction of the composite system to be $\gv{\Psi}=(\gv{\psi},\v{d})^{T}$, where the two components will be referred as s and d components 
     (which are still vectors) in the following. Then the Schr\"{o}dinger equation can be written as
    \begin{equation}
        \sum_{J}H_{IJ}\Psi_{J} = E \Psi_{I}.
    \end{equation}
    Here the indices $I,J$ can be continuous coordinates $\v{r}\in \mathbb{R}^3$ labeling the s components or be discrete $j\in \mathbb{Z}$ labeling the d components, 
    and the summation over $J$ combines the integration over continuous position and the summation over the discrete site index.
    Plugging Eq.\ (\ref{Hs}), (\ref{Hd}), (\ref{Ht}) into the Schr\"{o}dinger equation, we have
    \begin{gather}
        \left\{\left[\frac{(-i\partial_{\v{r}}+k_h\sigma_z\hat{x})^2}{2m}\right]\tau_{z}+\Delta\tau_{x}\right\} \psi(\v{r}) -t\sum_{j}\delta(\v{r}-\v{R}_j)\tau_z d_{j} = E \psi(\v{r}) \\
        -t\int dr^{3}\, \delta(\v{r}-\v{R}_i)\tau_z \psi({\v{r}}) + \sum_{j} \left\{\left[(\epsilon_{d}-\mu)\delta_{ij}-W_{ij}\right]\tau_{z}-K\v{S}\cdot\gv{\sigma}\delta_{ij}\right\}d_{j} = E d_{i}.
    \end{gather}
    The normalization condition is given by
    \begin{equation}
        \int d^3r\, \psi(r)^{\dagger} \psi(r) + \sum_{j} d^\dagger_j d_j = 1.
    \end{equation}
Notice that the s components $\psi(r)$ have units of [volume]$^{-1/2}$ while the d components are dimensionless. 

Now we turn to the Green function of the system, which reads
 \begin{equation}
     G(E)=\left( E-H \right)^{-1} = \left( E- H_0 -H_T \right)^{-1} = G_{0}(E) + G_{0}(E)H_TG_{0}(E) + G_{0}(E)H_TG_{0}(E)H_TG_{0}(E) + \cdots, \label{GE}
 \end{equation}
where $G_{0}(E)=(E-H_0)^{-1}$. Let us now define a reduced Green function $g$ in which we restrict the position arguments of the superconductor to the discrete impurity sites. For instance, in the $ss$ block, this reduced lattice Green function is defined as $g^{ss}_{ij}=G^{ss}(\v{R}_i,\v{R}_j)$ with $i,j\in \mathbb{Z}$ (as opposed to $G^{ss}(\v{R}_i,\v{R}_j)$ with $\v{R}_i,\v{R}_j \in \mathbb{R}^3$). Let's compute the separate blocks of $g$ in (s,d) space (space formed by superconductor and d levels of Anderson impurities),
\begin{equation}
    g_{ij}^{dd} = g_{0,ij}^{dd} + \sum_{k_1k_2} g_{0,ik_1}^{dd}(-t\tau_z)g^{ss}_{0,k_1k_2}(-t\tau_z)g_{0,k_2j}^{dd}+\cdots
\end{equation}
\begin{equation}
    g^{ss}_{ij} = g^{ss}_{0,ij} + \sum_{k_1k_2}g^{ss}_{0,ik_1}(-t\tau_z)g^{dd}_{0,k_1k_2}(-t\tau_z)g^{ss}_{0,k_2j}+\cdots
\end{equation}
\begin{equation}
    g^{ds}_{ij} = \sum_{k_1} g_{0,ik_1}^{dd}(-t\tau_z)g^{ss}_{0,k_1j} + \cdots.
\end{equation}
with the notation
\begin{equation}
    g_{0,ij}^{ss} = G^{ss}_0(\v{R}_i,\v{R}_j),\qquad 
    g_{0,ij}^{ds} = G^{ds}_0(i,\v{R}_j).
\end{equation}

We obtain the inverse of the lattice Green function, in matrix notation, 
\begin{equation}
    g^{-1} = \left(\begin{array}{cc}
        (g^{ss}_0)^{-1} & t\tau_{z}\\
        t\tau_{z} & E-H_{d}
    \end{array}\right), \label{eq.lattice_gf}
\end{equation}
leading to Eq.\ (5) of the main text.
Here, $g^{-1}$ is a matrix in (s,d) space, in site space, and in spin and particle-hole space.
(The spin and particle-hole indices are implicitly included in the site labels.) 

We first compute the Green function of the superconductor. For $i\neq j$, 
\begin{align*}
    g_{0,ij}^{ss}(E)&=\bra{\v{R}_i} (E-H_s)^{-1}\ket{\v{R}_j} \nonumber \\
    &= \bra{\v{R}_i} \left[E-\left( \frac{(\v{p}+k_{h}a\sigma_z\hat{x})^2}{2m}-\mu\right)\tau_z-\Delta\tau_x\right]^{-1} \ket{\v{R}_{j}} \nonumber \\
    &= \bra{\v{R}_i}e^{-ik_h x\sigma_z} 
    \left[E-\left( \frac{(\v{p}+k_{h}a\sigma_z\hat{x})^2}{2m}-\mu\right)\tau_z-\Delta\tau_x\right]^{-1} e^{ik_h x\sigma_z}\ket{\v{R}_{j}} \nonumber \\
    &=e^{-ik_{h}(i-j)a\sigma_z}\frac{1}{V}\sum_{\v{k}} \frac{e^{i\v{k}\cdot(\v{R}_i - \v{R}_j)}}{E-\xi_{k}\tau_z-\Delta\tau_x} \nonumber \\
    &=e^{-ik_{h}(i-j)a\sigma_z}\left[(E+\Delta\tau_x)P_{0}(\abs{i-j}a)+\tau_z P_{1}(\abs{i-j}a)\right]
\end{align*}
where \cite{pientka13}
\begin{gather}
    P_{0}(r)=\frac{\nu_{0}}{2}\int d\xi_{k}\int_{-1}^{1}dx\,\frac{e^{ikrx}}{E^2-\xi_{k}^{2}-\Delta^{2}}
    =-\frac{\pi\nu_{0}}{\sqrt{\Delta^{2}-E^{2}}}\frac{\sin k_{F}r}{k_{F}r}e^{-r/\xi_{E}}\\
    P_{1}(r)=\frac{\nu_{0}}{2}\lim_{\omega_D\rightarrow\infty}
    \int d\xi_{k}\int_{-1}^{1}dx\,\frac{\xi_{k}e^{ikrx}}{E^2-\xi_{k}^{2}-\Delta^{2}}\frac{\omega_{D}^{2}}{\xi_{k}^{2}+\omega_{D}^{2}}
    =-\pi\nu_{0}\frac{\cos k_{F}r}{k_{F}r}e^{-r/\xi_{E}}
\end{gather}
with $\nu_0$ the normal density of states at the Fermi energy and $\xi_{E}=v_{F}/\sqrt{\Delta^2-E^2}$. Then we obtain
\begin{equation}
    g_{0,ij}^{ss}(E) =-\pi\nu_{0} e^{-ik_{h}x_{ij}\sigma_z}\left\{ \frac{E+\Delta\tau_x}{\sqrt{\Delta^2-E^2}}\frac{\sin k_Fr_{ij}}{k_F r_{ij}}e^{-r_{ij}/\xi_{E}}
    +\tau_z \frac{\cos k_F r_{ij}}{k_F r_{ij}}e^{-r_{ij}/\xi_{E}}\right\}
\end{equation}
where $x_{ij}=x_{i}-x_{j}=(i-j)a$, $r_{ij}=\abs{x_{ij}}$. We can also rewrite it as
\begin{equation}
    g_{0,ij}^{ss}(E) =-\pi\nu_{0} e^{-ik_{h}x_{ij}\sigma_z}\left\{ \frac{E+\Delta\tau_x}{\sqrt{\Delta^2-E^2}}{\rm Im}{f(r_{ij})}
    +\tau_z{\rm Re}{f(r_{ij})} \right\}
\end{equation}
with \begin{equation}
    f(r) = \frac{e^{ik_Fr - r/\xi_{E}}}{k_F r}. 
\end{equation}
For $i=j$, we find (see Ref.\ \cite{pientka13} for details)
\begin{equation}
    g_{0,ii}^{ss}(E) = -\pi \nu_{0} \frac{E+\Delta\tau_x}{\sqrt{\Delta^2-E^2}}.
\end{equation}

\section{Green function in lattice momentum space}

We proceed by computing the Green function in lattice momentum space
\begin{equation}
    g^{ss}_0(k)=\sum_{j}e^{-ikx_{ij}}g^{ss}_{0,ij}=g^{ss}_{0,ii}+2\sum_{j=1}^{\infty}\cos((k+k_{h}\sigma_{z})ja)\tilde{g}^{ss}_{0j},
\end{equation}
where
\begin{equation}
    \tilde{g}^{ss}_{ij} =  - \pi\nu_0\frac{E+\Delta\tau_x}{\sqrt{\Delta^2-E^2}}{\rm Im}{f(r_{ij})}
    -\pi\nu_0\tau_z{\rm Re}{f(r_{ij})} .
\end{equation}
For convenience, we define
\begin{align}
F(k) & =-2\sum_{j=1}^{\infty}\cos(k_{F}ja)f(ja)\\
 & =\frac{1}{k_{F}a}\left[\ln(1-e^{-a/\xi_{E}+i(k_{F}+k)a})+\ln(1-e^{-a/\xi_{E}+i(k_{F}-k)a})\right],
\end{align}
which has the property $F(k)=F(-k)$, and
define $L^{\sigma_z}(k)=F(k+k_{h}\sigma_z)-i$.
We then obtain the Green function in lattice momentum space
\begin{equation}
    g^{ss}_0(k)=\frac{\pi\nu_{0}(E+\Delta\tau_{x})}{\sqrt{\Delta^{2}-E^{2}}}L^{\sigma_z}_{i}(k,E)+\pi\nu_{0}\tau_{z}L^{\sigma_z}_{r}(k,E)
\end{equation}
where $L^{\sigma_z}_{i}$ and $L^{\sigma_z}_r$ are the imaginary and real parts of $L^{\sigma_z}$, cf. Eq.\ (7) of the main text.
\begin{figure}
    \includegraphics[width=0.7\textwidth]{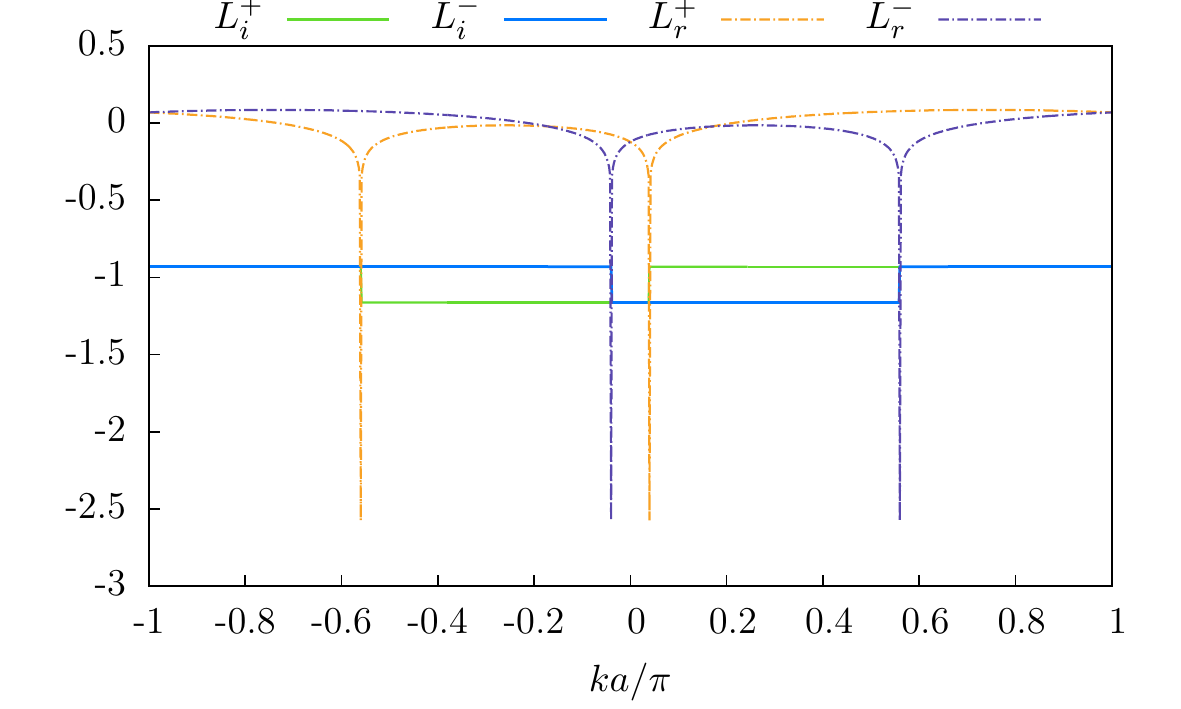}
   \caption{\label{LiLr}
    $L_i^\pm(k)$ and $L_r^\pm(k)$ for $k_F a=4.3\pi$, $k_h a=0.26\pi$ and $\xi_0/a=\infty$. }
\end{figure}

\section{Numerical calculation of physical quantities}
\subsection{Excitation spectrum}
The inverse Green function $g^{-1}$ is diagonal in momentum space and we have
\begin{align}
g^{-1}(k,E) & =\left(\begin{array}{cc}
    (g^{ss}_0)^{-1}(k,E) & t\tau_{z}\\
    t\tau_{z} & (g^{dd}_0)^{-1}(k,E) \end{array}\right),
\end{align}
with
\begin{equation}
(g^{dd}_0)^{-1}(k,E)=E-\left[\epsilon_d-2w\cos(k+k_h\sigma_z)a\right]\tau_z+KS\sigma_x.
\end{equation}
This is an $8\times8$ matrix as a function of $k\in[-\frac{\pi}{a},\frac{\pi}{a}]$.
Then the subgap band can be calculated by imposing the condition
\begin{equation}
    \det(g^{-1}(k,E))=0.
\end{equation}

\subsection{Majorana wavefunction}
Consider a finite chain of $N$ sites. Since the Majorana state $\ket{\psi^{M}}\in\ker g^{-1}(E=0)$,
its real space representation in terms of a column vector $\psi^{M}(i)=\braket{i|\psi^{M}}$,
with $i=1,\dots N$, is in the kernel of the $8N\times 8N$ matrix $g_{ij}^{-1}$ evaluated at $E=0$.
The occupation probability at site $i$ is
given by
\begin{equation}
    \abs{\psi^{M}(i)}^{2}=\langle\psi^{M}(i),\psi^{M}(i)\rangle,
\end{equation}
where $\langle\cdot,\cdot\rangle$ denotes the inner product in spin and particle-hole space. We then
take either the components of the superconducting host when $\Gamma>\Delta$ or take the d components
when $\Gamma<\Delta$. Note that the s and d entries have different units and cannot be added. 
However, the localization length of the end state is the same for both components. 

\subsection{Local density of states}

The local density of states is related to the diagonal elements of
\begin{equation}
    A^{\mu\nu}(\v{r},E)=-\frac{1}{\pi}\lim_{\eta\rightarrow 0^+}{\rm Im}\Tr_{\sigma}G^{\mu\sigma,\nu\sigma}(E+i\eta,\v{r},\v{r}),
\end{equation}
where $\mu,\nu \in\{s,d\}\times\{e,h\}$ are composite indices for (s,d) and (e,h) (particle-hole) components, 
and $\sigma$ is the index of the spin space, within which the trace is taken. To obtain a true tunneling density of states, we would also need to use a Green function $G$ which includes information about the spatial structure of the adatom d-states. This structure is not explicit in the Green function defined in Eq.\ (\ref{GE}). The latter treats the adatoms in tight-binding approximation and retains only the amplitudes of the atomic d-orbitals rather than the spatial structure of the corresponding wavefunctions.

In the case of $\Gamma\gg\Delta$, there is a strong transfer of the spectral weight of the subgap excitations to the superconductor. 
Therefore, apart from a narrow vicinity of the adatoms, the main contribution to the local density of states comes from the states of the host. In this limit, we neglect the contribution of the adatoms (i.e. the d-levels) to the local density of states. Focusing on the electron contribution, we compute $A^{\mu\nu}(\v{r},E)$ for $\mu=\nu=(s,e)$ and $\v{r}=\v{R}_j$ at site $j$. Explicitly, we are able to find the local density of states at each site $j$, 
\begin{equation}
    A_{j}(E) = A^{\mu\mu}(\v{R}_j,E)=-\frac{1}{\pi}\lim_{\eta\rightarrow 0^+}{\rm Im}\Tr_{\sigma}g^{\mu\sigma,\mu\sigma}_{jj}(E+i\eta), \qquad \mu=(s,e),
\end{equation}
from the lattice Green function $g$ defined in Eq.\ (\ref{eq.lattice_gf}). It should also be mentioned that we include a finite imaginary part in the energy $E\rightarrow E+i\eta$, with $\eta=0.015\Delta$. This small $\eta$ put in by hand introduces a finite broadening of the $\delta$-peaks which we obtain from a finite size calculation. This results in a smooth local density of states.

\section{Derivation of Equation (10) in the Main text}

The dressed Green function of the superconductor including the self-energy from hybridization with the magnetic adatoms
can be written as
\begin{equation}
    g^{ss}=g^{ss}_0(1-\Sigma g^{ss}_0).
\end{equation}
The self-energy takes the form
\begin{align}
    \Sigma &=t^2 g^{dd}_0=\left(\begin{array}{cccc}
        E-\varepsilon_{d}^{+} & KS & 0 & 0\\
        KS & E-\varepsilon_{d}^{-} & 0 & 0\\
        0 & 0 & E+\varepsilon_{d}^{+} & KS\\
        0 & 0 & KS & E+\varepsilon_{d}^{-}
    \end{array}\right)^{-1} \nonumber \\
    &=\frac{t^2}{(E-\varepsilon_{d}^{+})(E-\varepsilon_d^-)-(KS)^2}
    \left(\begin{array}{cccc}
        E-\varepsilon_{d}^{-} & -KS & 0 & 0\\
        -KS & E-\varepsilon_{d}^{+} & 0 & 0\\
        0 & 0 & 0 & 0\\
        0 & 0 & 0 & 0
    \end{array}\right)
    +\frac{t^2}{(E+\varepsilon_{d}^{+})(E+\varepsilon_d^-)-(KS)^2}
    \left(\begin{array}{cccc}
        0 & 0 & 0 & 0\\
        0 & 0 & 0 & 0\\
        0 & 0 & E+\varepsilon_{d}^{-} & -KS\\
        0 & 0 & -KS & E+\varepsilon_{d}^{+}
    \end{array}\right) 
\end{align}
where $\varepsilon_d^{\pm}(k)=\epsilon_d -2w\cos(k\pm k_h)a$. We now assume that the spin-splitting $KS$ is the largest energy scale in the problem and is taken to $\infty$. In this limit, the spin direction is frozen along the Zeeman axis. Moreover, we take the limit such that the energy of the spin-up band goes to $-\infty$ while the energy $E_d$ of the spin-down band remains finite. To take the limit, we temporarily introduce $E_{d}^{\pm}=\varepsilon_d^\pm + KS$, and replace $\varepsilon_d^\pm = E_d^\pm -KS$ in the self-energy. Now, $E$ and $E_d^\pm$ remain finite in the limit $KS\to\infty$, i.e., $KS\gg E, E_{d}$, and we can approximate
\begin{subequations}
\begin{gather}
    (E-\varepsilon_d^{+})(E-\varepsilon_d^-)-(KS)^2 \simeq -KS(E_d^++E_d^- -2E) \\
    (E+\varepsilon_d^{+})(E+\varepsilon_d^-)-(KS)^2 \simeq -KS(E_d^++E_d^- +2E) 
\end{gather}
\end{subequations}
in the denominators and 
\begin{equation}
    E\pm \varepsilon_d^{+,-} \simeq \mp KS 
\end{equation}
in the matrix elements. Thus, for $KS\rightarrow \infty$, we find
\begin{equation}
    \Sigma \simeq  \frac{\alpha_-}{2\pi\nu_0}(1+\tau_z)(\sigma_x-\tau_z)
    +\frac{\alpha_+}{2\pi\nu_0}(1-\tau_z)(\sigma_x-\tau_z),
\end{equation}
where 
\begin{equation}
    \alpha_{\pm} = \frac{\pi\nu_0 t^2}{2E_d \pm2E}, \qquad E_{d}=\frac{E_{d}^{+}+E_{d}^{-}}{2}.
\end{equation}
Note that we can also write
\begin{equation}
    \Sigma =e^{-i\frac{\pi}{4}\sigma_y}
    \left\{\frac{\alpha_-}{2\pi\nu_0}(1+\tau_z)(\sigma_z-\tau_z)
    +\frac{\alpha_+}{2\pi\nu_0}(1-\tau_z)(\sigma_z-\tau_z)\right\}
    e^{-i\frac{\pi}{4}\sigma_y}
        = e^{i\frac{\pi}{4}\sigma_y}
    \left(\begin{array}{cccc}
        0 & 0 & 0 & 0\\
        0 & -\frac{2\alpha_{-}}{\pi\nu_{0}} & 0 & 0\\
        0 & 0 & \frac{2\alpha_{+}}{\pi\nu_{0}} & 0\\
        0 & 0 & 0 & 0
    \end{array}\right)
    e^{-i\frac{\pi}{4}\sigma_y}.
\end{equation}
Then
\begin{align}
    \det(1-\Sigma g^{ss}_0) &= \det\left\{1- 
    \left(\begin{array}{cccc}
        0 & 0 & 0 & 0\\
        0 & -\frac{2\alpha_{-}}{\pi\nu_{0}} & 0 & 0\\
        0 & 0 & \frac{2\alpha_{+}}{\pi\nu_{0}} & 0\\
        0 & 0 & 0 & 0
    \end{array}\right)
  \left(\begin{array}{cccc}
        \mathcal{A}+\mathcal{B} & \mathcal{E}+\mathcal{F} & \mathcal{C} & \mathcal{D}\\
        \mathcal{E}+\mathcal{F} & \mathcal{A}+\mathcal{B} & \mathcal{D} & \mathcal{C}\\
        \mathcal{C} & \mathcal{D} & \mathcal{A}-\mathcal{B} & -\mathcal{E}+\mathcal{F}\\
        \mathcal{D} & \mathcal{C} & -\mathcal{E}+\mathcal{F} & \mathcal{A}-\mathcal{B}
    \end{array}\right)\right\} \nonumber \\
    &=\det\left\{ 1-2 
        \left(\begin{array}{cc}
            -\alpha_{-}(\mathcal{A}+\mathcal{B}) & -\alpha_{-}\mathcal{D}\\
            \alpha_{+}\mathcal{D} & \alpha_{+}(\mathcal{A}-\mathcal{B})
        \end{array}\right)
    \right\} \nonumber \\
   &=\left[ 1+2\alpha_{-}(\mathcal{A}+\mathcal{B}) \right]\left[ 1-2\alpha_+(\mathcal{A}-\mathcal{B}) \right]+4\alpha_+\alpha_-\mathcal{D}^2 \nonumber \\
    &= 1+2\mathcal{A}(\alpha_- -\alpha_+) +2\mathcal{B}(\alpha_- + \alpha_+) +4\alpha_+ \alpha_- (\mathcal{D}^2+\mathcal{B}^2-\mathcal{A}^2)
\end{align}
where
\begin{gather}
    \mathcal{A}=\frac{E}{\sqrt{\Delta^{2}-E^{2}}}L_i\qquad \mathcal{B}=L_r\qquad \mathcal{C}=\frac{\Delta}{\sqrt{\Delta^{2}-E^{2}}}L_i\\
    \mathcal{D}=-\frac{\Delta}{\sqrt{\Delta^{2}-E^{2}}}(\delta L_i)  \qquad \mathcal{E}=-(\delta L_r)\qquad \mathcal{F}=-\frac{E}{\sqrt{\Delta^{2}-E^{2}}}(\delta L_i)   
\end{gather}
and
\begin{equation}
    L_{i,r} = \frac{L_{i,r}^{+}+L_{i,r}^{-}}{2} \qquad    
    (\delta L_{i,r}) = \frac{L_{i,r}^{+}-L_{i,r}^{-}}{2}.
\end{equation}
The excitation spectrum can be obtained by requiring the above determinant to be zero,
\begin{align}
    0 &=1+\frac{2E}{\sqrt{\Delta^{2}-E^{2}}}L_{i}(\alpha_{-}-\alpha_{+})+2L_{r}(\alpha_{-}+\alpha_{+})+4\alpha_{+}\alpha_{-}
    \left\{\frac{\Delta^{2}}{\Delta^{2}-E^{2}}(\delta L_{i})^{2}-\frac{E^{2}}{\Delta^{2}-E^{2}}L_{i}^{2}+L_{r}^{2}\right\} \nonumber \\
    &=1+\frac{2\Gamma E_{d}}{E_{d}^{2}-E^{2}}L_{r}+\frac{\Gamma^{2}}{E_{d}^{2}-E^{2}}L_{r}^{2} \nonumber \\
    &+\frac{2E^{2}}{\sqrt{\Delta^{2}-E^{2}}}\frac{\Gamma}{E_{d}^{2}-E^{2}}L_{i}
    +\frac{\Gamma^{2}}{E_{d}^{2}-E^{2}}\left\{\frac{\Delta^{2}}{\Delta^{2}-E^{2}}(\delta L_{i})^{2}-\frac{E^{2}}{\Delta^{2}-E^{2}}L_{i}^{2}\right\}
\end{align}
with $\Gamma=\pi\nu_0t^2$. Multiplying by $(\Delta^2-E^2)(E_d^2  -E^2)$ gives Eq.\ (9) of the main text,
\begin{equation}
      (\Delta^2-E^2)[E_d+\Gamma L_r]^2  
          - E^2 [\sqrt{\Delta^2-E^2}-\Gamma L_i]^2 + \Gamma^2 \Delta^2 (\delta L_i)^2 = 0.
\end{equation}

\section{Comparison with experiment}

Here, we briefly discuss the Majorana localization length as given in Eq.\ (2) of the main text in more detail. For lead, the literature value of the coherence length $\xi_0$ is $\xi_0\simeq 80$nm \cite{Fetter}. 

Given that the Fe adatoms are directly embedded into the host superconductor, the hybridization $\Gamma$ is governed by physics on the atomic scale. Thus, $\Gamma$ should also be of the order of atomic energies, $\Gamma\sim 1$eV (for a similar estimate of the hybridization, see Ref.\ \cite{li14}). According to this estimate, the adatom chains are deeply in the regime of strong hybridization, $\Gamma\gg\Delta$, with quasiparticle weight $Z \simeq \Delta/\Gamma \sim 10^{-3}$. 

The value of $\Gamma$ assumed in this estimate is comparable to $\hbar v_F/a$, and thus at the border of applicability of our theory.

The ratio of gaps $\alpha = \Delta_{\rm top}/\Delta$ is controlled by the strength of spin-orbit coupling in the substrate superconductor. This is difficult to estimate from first principles. Extracting $\Delta_{\rm top}$ from the experiment [5] is hindered by the substantial broadening of the peaks in the STM spectra. Interpreting these maxima as van-Hove singularities of the Shiba bands, one finds a topological gap of order $0.1\Delta$. 

Inserting these numbers into Eq.\ (2) for the Majorana localization length $\xi_M$, we find that $\xi_M$ is two orders of magnitude smaller than the coherence length $\xi_0$ of the substrate superconductor, i.e., of order 1nm or a few adatom spacings $a$ ($a\simeq 0.4$nm according to Ref.\ \cite{ali}). 
\end{widetext}

\begin{thebibliography}{99}
\bibitem{L} The function $L$ is plotted in \cite{supp}. Roughly, $L^{\sigma_z}_i$ (and hence $L_i$) is of order unity. $L^{\sigma_z}_r$ (and hence $L_r$) is of order $1/k_Fa$, except at isolated values of $k$ where it has logarithmic singularities. $\delta L_i$ should be thought of as a dimensionless measure of the effective SO strength, which is also of order $1/k_Fa$ for our spin-helix model.
\end{thebibliography}
\end{document}